# Analysis of fragment yield ratios in the nuclear phase transition

R. Tripathi[1,*,$], A. Bonasera[1,2], S. Wuenschel[1,3], L. W. May[1,3], Z. Kohley[1,3], G. A. Souliotis[1,4], S. Galanopoulos[1,#], K. Hagel[1], D. V. Shetty[1,##], K. Huseman[1], S. N. Soisson[1,3], B. C. Stein[1,3], and S. J. Yennello[1,3]

1. Cyclotron Institute, Texas A&M University, College Station, 77843, TX, USA
2. Laboratori Nazionali del Sud-INFN, v. S. Sofia 64, 95123 Catania, Italy
3. Chemistry Department, Texas A&M University, College Station, 77843, TX, USA
4. Laboratory of Physical Chemistry, Department of Chemistry, National and Kapodistrian University of Athens, Athens, Greece

**Abstract**

The critical phenomena of the liquid-gas phase transition has been investigated in the reactions $^{78,86}$Kr+$^{58,64}$Ni at beam energy of 35 MeV/nucleon using the Landau free energy approach with isospin asymmetry as an order parameter. Fits to the free energy of fragments showed three minima suggesting the system to be in the regime of a first order phase transition. The relation $m = -\partial F/\partial H$, which defines the order parameter and its conjugate field $H$, has been experimentally verified from the linear dependence of the mirror nuclei yield ratio data, on the isospin asymmetry of the source. The slope parameter, which is a measure of the distance from a critical temperature, showed a systematic decrease with increasing excitation energy of the source. Within the framework of the Landau free energy approach, isoscaling provided similar results as obtained from the analysis of mirror nuclei yield ratio data. We show that the external field is primarily related to the minimum of the free energy, which implies a modification of the source concentration $\Delta$ used in isospin studies.

* email: rtripathi@comp.tamu.edu

$ On leave from Radiochemistry Division, Bhabha Atomic Research Centre, Mumbai, India.

# Present address: Hellenic Army Academy, Department of Physical Sciences & Applications, Athens, Greece

## Present address: Physics Department, Western Michigan University, Kalamazoo, 49008, MI, USA

Investigation of the nuclear liquid-gas phase transition is currently one of the important research objectives of heavy-ion collisions in the Fermi energy domain. Various signatures have been employed to investigate the critical phenomena in nuclear systems [1-11]. Recently, Bonasera et al. [11,12] used fragment yield data from different reactions to investigate the nuclear phase transition using the Landau free energy approach [13,14], which is applicable to the systems in the vicinity of a critical point. In this work, isospin degree of freedom was identified as an additional order parameter in nuclear phase transition. In this approach, the free energy per nucleon $F$ of a fragment is related to an order parameter $m$ as given by the following equation

$$\frac{F}{T} = \frac{1}{2}am^2 + \frac{1}{4}bm^4 + \frac{1}{6}cm^6 - \frac{H}{T}m \qquad (1)$$

where $m = (N-Z)/A$, $N$, $Z$ and $A$ are the neutron, proton and mass numbers of the fragment respectively. The quantity $m$, which is a measure of the isospin asymmetry of the fragment, can be defined as an order parameter if $m = -\partial F/\partial H$, where $H$ is its conjugate variable [13,14]. The coefficients $a$, $b$ and $c$ are fitting parameters [11,13] and $T$ is the temperature of the fragmenting source. In absence of any external field i.e. $H/T=0$, Eq (1) may predict three minima corresponding to $\pm m_+$ and $m_0$ [12,13], where

$$\pm m_+ = \pm \frac{-b + \sqrt{b^2 - 4ac}}{2c}$$
$$m_0 = 0 \qquad (2)$$

The presence of an external field shifts the positions of these minima. Differentiating Eq (1) and substituting $(m'+\varepsilon')$ as a general solution in presence of an external field gives [12]

$$a(m' + \varepsilon') + b(m' + \varepsilon')^3 + c(m' + \varepsilon')^5 - \frac{H}{T} = 0 \qquad (3)$$

Solving Eq (3) for small $\varepsilon$, gives new positions of the minima as ($\pm m+\varepsilon$) and $\varepsilon_0$. Shifts in the minima positions are given by

$$\varepsilon_0 = \frac{H/T}{a}$$

$$\varepsilon = \frac{-H/T}{\left[4a + \frac{-b^2 + b\sqrt{b^2 - 4ac}}{c}\right]} \qquad (4)$$

Imposing the condition for a first order phase transition $b = -4\sqrt{ac/3}$ gives $\varepsilon = (H/T)/4a$. The coefficient 'a' is related to the temperature of the system relative to a critical temperature [13]. These solutions can be tested from the fits to the experimental free energies.

Information about the coefficient 'a' and $H/T$ can also be obtained from dependence of the appropriate yield ratios on the isospin asymmetry of the source. Based on a modified Fisher model [5,11], fragment yields are proportional to $A^{-\tau}e^{-(F/T)A}$, where $A$ is the fragment mass number and $\tau$ is the critical exponent. In an earlier work [11], the critical exponent $\tau$ was determined as 2.3 from the power law dependence of mass yields, which is a signature of critical behavior. Thus, using Eq (1) and (4), it can be shown that, for a pair of mirror nuclei or for fragments of a given type arising from sources with different isospin asymmetry ($m_s$), the power law dependence cancels out exactly and the ratio of yields is given as

$$\frac{1}{A}\ln\left(\frac{Y_2}{Y_1}\right) = -\frac{F_2 - F_1}{T} A = \frac{\Delta F}{T}$$
$$= m_2 \frac{H_2}{T} - m_1 \frac{H_1}{T} \qquad (5)$$

Substituting $H/T = a\varepsilon_0$ from Eq (4) and assuming that $\varepsilon_0$ is proportional to the isospin asymmetry $m_s$ of the source ($\varepsilon_0 = (a'/a)m_s$) gives

$$\frac{1}{A}\ln\left(\frac{Y_2}{Y_1}\right) = m_2(a\varepsilon_{0,2}) - m_1(a\varepsilon_{0,1})$$
$$= a'(m_2 m_{s2} - m_1 m_{s1}) \qquad (6)$$

Thus, Eq (6) can be used to explain the fragment yields in terms of their dependence on the isospin asymmetry of the fragmenting source and extract information about the slope parameter which is related to the coefficient '$a$'. We stress that, based on isoscaling results [26,27] we would expect that $m_s = \varepsilon_0$, which is not true as we will show below. This can be tested experimentally by determining the mirror nuclei yield ratios in the fragmentation of sources spanning a wide range of $m_s$ values. This requirement can be fulfilled by studying the fragmentation of the projectile like source (quasiprojectile) formed in peripheral and mid-peripheral collisions. With improved 4π-multi-detector systems it is possible to reconstruct such events by measuring the charge, mass and momentum of the detected particles/fragments. Reconstruction of the quasiprojectile leads to a better characterization of events and, in turn, a better control over the $m_s$ value. Furthermore, thermodynamic properties (such as the temperature) of the fragmenting source can also be determined [18]. The linear dependence predicted by Eq (6) for the yield ratio of a mirror nuclei pair is also required for '$m$' to be an order parameter.

In the present work, fragment yield data from the quasiprojectile fragmentation in the reactions $^{78,86}$Kr+$^{58,64}$Ni at beam energy of 35MeV/nucleon have been analyzed using Landau free energy approach. A detailed analysis of the yield ratios of mirror nuclei pairs for $A=3$ ($^3$H, $^3$He) and $A=7$ ($^7$Li, $^7$Be) has been carried out to test the predictions from Landau free energy approach and extract the slope parameter. Variation of the slope parameter with excitation energy of the source has also been investigated. An analysis, analogous to the conventional isoscaling and $m$-scaling [12] has also been carried out within the framework of Landau free energy approach. These isotopes were chosen because of the available statistics over a large range of the isospin asymmetry of the source. Moreover, with increasing $Z$ of the fragment, Coulomb effects may become significant and complicate the analysis. This analysis is important in the context of the Landau's approach since from mirror nuclei and isoscaling we could fix $H/T$ entering Eq (1) and eventually, using Eq (4), also the parameter '$a$' could be fixed. This will result in more constraints in Eq (1) to reproduce the experimental free energy.

The experiments were performed at the Texas A&M University K500 superconducting cyclotron. Charged particles were detected using the NIMROD-ISiS array [15,16]. Neutrons were detected with the TAMU neutron ball [15] surrounding the NIMROD-ISiS array. The details of the experiment can be found in [17,18]. The quasiprojectile source was reconstructed by selecting events with the condition that the longitudinal velocity of the fragments with Z=1, 2, ≥3 be, respectively, in the range of ±65%, 60% and 40% of the velocity of the heaviest fragment in the event [18]. Further, the total $Z$ of the detected fragments was selected to be in range $Z$=30-40 encompassing the projectile $Z$ of 36. Using the four reaction systems, the yield ratios of mirror nuclei were determined over a wide range of $m_s$ from -0.03 to 0.21. The $m_s$ values were calculated after correcting for free neutrons emitted by the quasiprojectile [17,18].

The data on fragment yields were divided into four $m_s$ bins 0±0.03, 0.06±0.03, 0.12±0.03 and 0.18±0.03 with mean $m_s$ values at 0.010, 0.062, 0.115 and 0.169 respectively. In further discussion, these $m_s$ bins will be referred to with their mean values. A typical plot of fragment yields for $m_s$=0.169 is shown in Fig. 1(a) (bottom panel). As seen from the figure, there is no systematic trend in the data. Free energies obtained by normalizing the fragment yields with respect to $^{12}$C yield as discussed in ref [11] are shown in Fig. 1(b) (top panel). In the normalization procedure, $\tau$=2.3±0.1 from ref [11] was used. It should be mentioned that only the data for fragments with $m \neq 0$ (except $^{12}$C) are shown to exclude pairing effects in the analysis, which is particularly significant for lighter fragments. Investigations on the pairing effects are currently going on and the results will be communicated in another paper. It can be seen from Fig. 1 (b) that the free energy shows a minimum close to, but not exactly at, $m$=0. The uncertainty on the data points includes statistical error, uncertainty on $\tau$ and an additional 10% systematic error. The dashed line in Fig. 1(a) is a fit to the data using Eq (1) with '$a$' and '$H/T$' as free parameters ($b=c=0$), as might be suggested from the symmetry energy entering the Weizacker mass formula. The solid line is a fit using the complete Eq (1) with $a$, $b$, $c$ and $H/T$ as free parameters. It can be seen from this figure that the complete Landau equation provides a better fit to the free energy data. This was found to be true for the data of other $m_s$ bins also. The average values of the coefficients $a$, $b$ and $c$ were obtained as 18.2±2.8, -120±37 and 138±50 respectively. The uncertainty on the parameters is the standard deviation of their values over different $m_s$ bins. The scatter in the values of $a$, $b$ and $c$, as reflected from the large standard deviation, was mainly due to the large number of parameters, absence of data points at large '$m$' values except for protons and neutrons and correlation of '$H/T$' and '$a$' as evident from Eq (4). Therefore, in order to better constrain the fit, $H/T$ values were fixed from mirror nuclei yield ratios as $H/T = 0.5\ln(Y_2/Y_1)$, where $Y_2$ and $Y_1$ are respectively the yields of the neutron rich and neutron poor members of the mirror nuclei pair. Using the mirror nuclei yield ratio data for $A=3$ and 7, the $H/T$ values corresponding to the $m_s$ values of 0.010, 0.062, 0.115 and 0.169 were obtained as 0.026±0.081, 0.390±0.095, 0.768±0.104 and 1.120±0.048 respectively. After fixing $H/T$, the average values of $a$, $b$ and $c$ were obtained as 15.1±0.5, -89±5 and 101±7, showing a large reduction in the scatter of the parameter values over different $m_s$ bins. Thus, $H/T$ values obtained from the mirror nuclei yield ratio data helped better constrain the fit using Eq (1). It should be mentioned here that the coefficient $a$, $b$ and $c$ satisfy the condition for a first order phase transition $b = -4\sqrt{ac/3}$ [13] within the error bars. The values of $\varepsilon_0$ calculated using Eq (4) were 0.002±0.005, 0.025±0.006, 0.053±0.008 and 0.075±0.006 for $m_s$=0.010, 0.062, 0.115 and 0.169 respectively. These $\varepsilon_0$ values were in excellent agreement with the position of the central minima in the free energy plot.

In order to understand the relation between $\varepsilon_0$ and $m_s$, mirror nuclei yield ratio data were analyzed in the light of Eq (6). For a pair of mirror nuclei arising from a source of isospin asymmetry $m_s$, their yield ratio can be written as

$$\frac{1}{2}\ln\left(\frac{Y_2}{Y_1}\right) = a'm_s \qquad (7)$$

Fig. 2 (a) shows a plot of '0.5 ln($Y_2/Y_1$)' for the four reaction systems as a function of mean $m_s$ values. The error bars on the data are statistical errors. It can be seen in Fig. 2(a) that the mirror nuclei yield ratios show a linear dependence, as predicted by Eq (7), for all four reaction systems. It should be mentioned here that the number of $m_s$ bins have been increased by reducing the bin size to ±0.01. This provided more number of data points and thus better estimate of the slope parameter. Linear fitting to the yield ratio data from different reaction systems gave slope values in close agreement as shown in Table 1. This observation suggests that once the experimental data are selected with a specific $m_s$ value of the fragmenting source, they become independent of the reaction system. To further confirm this aspect, ordinate values from Fig. 2(a) for different reaction systems were averaged and subjected to linear fitting, as shown in Fig. 2(b). Filled and open symbols correspond to $A$=3 and 7 respectively. The slope values obtained from the linear fitting of the average values were also in close agreement with those obtained from the individual reaction systems, as shown in Table 1. The observed linearity in Fig. 2 indicates that the condition $m=-\partial F/\partial H$ is fulfilled and $m$ is an order parameter [13]. It can be seen from Table 1 that the slope values for $A$=3 and 7 are in reasonable agreement, suggesting that Coulomb effects are small, and the symmetry energy is the dominant contribution to the free energy. Since we are considering odd $A$ nuclei, pairing might be neglected if evaporation effects are not important, i.e. if the yields at freeze-out are not strongly modified due to secondary decays. The linear dependence of mirror nuclei yield ratio data on isospin asymmetry of the source is also expected based on the grand canonical calculations as shown in ref [19]. However, the present studies reveal that the proportionality constant $a'$ (Table 1) is different from the '$a$' value obtained by fitting the free energy data in Fig. 1(a). Using the value of $a'$ as 6.9 from Table I, we get $a'/a$ as 0.457±0.025, suggesting that $\varepsilon_0=0.457m_s$. In order to confirm this relation, $\varepsilon_0$ values, obtained from fits to the free energy, are plotted as a function of $m_s$ in Fig. 3. A liner fit to this plot gives slope value as 0.465±0.047, which is close the value obtained as $a'/a$. The lower value of $\varepsilon_0$ compared to $m_s$ suggests lower average isospin asymmetry of fragments ($<m_f>$) compared to $m_s$. For comparison, $<m_f>$ values calculated with and without neutrons and protons are also shown in Fig. 3. It can be seen from this figure that the average isospin asymmetry values, calculated with neutrons and protons, significantly deviate from the $\varepsilon_0$ values, which may be due to the large $m$ value for neutrons ($m=+1$) and protons ($m=-1$). Whereas, $<m_f>$ values calculated without neutrons and protons are in good agreement with the respective $\varepsilon_0$ values. This agreement can be understood from the fact that the value of $\varepsilon_0$ i.e. position of central minimum in Fig. 1(a) is mainly constrained by the heavier fragments and neutrons and protons may have only little effect (they play a larger role for the position of the other minima of the Landau's free energy). The lower average isospin of the fragments compared to the source may be driven by the larger mixing entropy for lower isospin asymmetry. This observation suggests that the plots similar to those in Fig. 2 as a function of $<m_f>$ corresponding to respective $m_s$ bins, calculated without neutrons and protons, should give slope value in agreement with the coefficient '$a$', which was indeed the case. This method gave slope values as 16.7±0.4 and 16.3±0.8 for $A=3$ and 7 respectively which are in reasonable agreement with the '$a$' value obtained from the fits to the free energy using Eq (1). The slightly larger value of the slope parameter may be due to the Coulomb effect at low $m_s$ value, which will shift the $<m_f>$ to comparatively larger values. The agreement between the slope parameter and the coefficient '$a$' suggests that, though, the ratio $a''/a$ i.e. the relation between $\varepsilon_0$ and $m_s$ may vary from system to system, mirror nuclei data can be used to extract the coefficient $a$ and $H/T$ directly by plotting against $<m_f>$ calculated without neutrons and protons. These values could be used to reduce the number of parameters in Eq (1). A fit to the data using Eq (1) with '$H/T$' and '$a$' obtained from mirror nuclei yield ratio data is shown as 'Fit_3' in Fig. 1(a).

The coefficient 'a' or the slope parameter $a'$ is a measure of the temperature of the system relative to a critical temperature. Therefore, its variation with excitation energy of the source was investigated. In order to carry out this study, mirror nuclei yield ratio data for each $m_s$ bin was further divided into excitation energy bins of 0.6 MeV. Based on the fact that the mirror nuclei yield ratios become independent of the reaction system after sorting the data according to $m_s$, the data of all the four reaction systems were combined to improve the statistics. It was observed that the sensitivity of the mirror nuclei yield ratio to the isospin asymmetry of the source decreased with increasing excitation energy, an observation similar to that in ref [20,21]. Fig. 4 shows a plot of the slope parameter $a'$ as a function of excitation energy for $A=3$ and 7. It can be seen from this figure that $a'$ values systematically decrease with increasing excitation energy of the fragmenting source suggesting that the temperature of the system is approaching closer to a critical temperature.

In the literature [11,12], a rough physical interpretation of the slope parameter 'a' can be obtained from the equivalence of the quantity $F/T$ with the symmetry energy per nucleon normalized with respect to the temperature (this indeed neglects entropy effects which might be important). Ignoring the higher order terms, the coefficient '$a/2$' of the first term in Eq (1) can be equated to $C_{sym}/T$, where $C_{sym}$ is the symmetry energy coefficient obtained from conventional isoscaling studies [17-18,22-29]. Within the framework of Landau free energy approach, isoscaling was carried out by taking the ratio of yields of the same fragments arising from two different sources with different $m_s$ values as done in the conventional isoscaling [12, 17-18, 22-29]. In this case Eq (6) reduces to

$$\frac{1}{A}\ln\left(\frac{Y_{ms2}}{Y_{ms1}}\right) = a'm(m_{s2} - m_{s1}) \qquad (8)$$

where $Y_{ms2}$ and $Y_{ms1}$ are, respectively, the yields of a given fragment with mass $A$ from fragmenting sources with isospin asymmetry values $m_{s1}$ and $m_{s2}$. The relation between the coefficient 'a' with $C_{Sym}/T$ makes Eq (8) equivalent to that derived from grand canonical calculations [26], provided that, at variance with previous assumptions, the source isospin asymmetry in Eq (8) is replaced by average isospin asymmetry of fragments calculated without neutrons and protons. The yield ratios for $^3$H, $^3$He, $^7$Li and $^7$Be were calculated for various possible combinations of $m_s$ bins such that $m_{s2} > m_{s1}$ to generate a plot according to Eq (8). For each $m_s$ bin, the yield of a given fragment was normalized with respect to the total number of events in that bin before taking the ratio. As expected from Eq (8), a reasonably good linearity (suggesting $m$ as an order parameter and $H$ its conjugate field) in the plot of '$(1/A)\ln(Y_{ms2}/Y_{ms1})$' as a function of '$m(m_{s2}-m_{s1})$' can be seen in Fig. 5. The slope ($a'$) of this plot is 6.82±0.10, which is in good agreement with the slope value obtained from the mirror nuclei yield ratios. The $a'$ value of 6.82 gives $a=a'/0.457=14.9$. Using $a=14.9$ and $C_{Sym}$ value of 25 MeV, the temperature of the system is obtained as 3.3 MeV. The temperature value appears to be on the lower side suggesting the requirement of further investigation on the relation between '$a$' and $C_{Sym}$, which will be done in a following paper. The slope values were also determined from the isoscaling plots for $E^*/A=4.6, 5.2$ and $5.8$ MeV/nucleon. This excitation energy range was chosen due to the larger statistics of the data. The slope values obtained from the isoscaling plots at different excitation energies (open squares in Fig. 4) were in reasonable agreement with those obtained from the analysis of the data of mirror nuclei yield ratios.

To conclude, fragment yield data have been analyzed using the Landau free energy approach, with isospin asymmetry as an order parameter. The Landau equation successfully explained the free energy of fragments arising from sources with different isospin asymmetry. Fixing the external field from the mirror nuclei yield ratio data provided a better constrain on the fit. Mirror nuclei yield ratio data showed a linear dependence on the isospin asymmetry ($m_s$) of the source, as expected in the Landau approach, suggesting isospin asymmetry '$m$' to be an order parameter. The dependence of *H/T* on average fragment isospin asymmetry (Eq 4) suggests that the mirror nuclei yield ratio or the isoscaling parameter primarily depends on the average isospin asymmetry of fragments, which, in turn, depends on the isospin asymmetry of the fragmenting source. The slope parameter, which is related to the temperature of the system relative to a critical temperature, showed a systematic decrease with increasing excitation energy of the source. Present studies showed that the difference between the slope parameter obtained from mirror nuclei yield ratio data and the coefficient '$a$' of Landau equation can be attributed to the difference in source isospin asymmetry and average fragment isospin asymmetry. Within some approximation, the coefficient '$a$' can also be related to the ratio of symmetry energy and temperature, obtained from conventional isoscaling studies. In analogy with the conventional isoscaling, yield ratios of similar fragments arising from sources with different isospin asymmetry ($m_s$) were plotted against an appropriate quantity for the abscissa which was a function of the isospin asymmetry of the fragment and the source, similar to the *m*-scaling proposed in [12]. The slope parameter obtained from this plot was in reasonable agreement with that obtained by fitting the mirror nuclei yield ratios.

Thus, a detailed analysis of the data on fragment yields within the framework of Landau free energy approach showed signature of a first order phase transition with respect to isospin degree of freedom. The results of these different analyses were observed to be mutually consistent. Comparison of the present results with existing fragmentation models and further investigation of the relationship between $a$ and $C_{sym}$ will be the objective of our future work.

This work was supported by the U.S. DOE grant DE-FG03-93ER40773 and the Robert A. Welch Foundation grant A-1266.

Table 1. Slope parameter (*a′*) obtained by fitting the mirror nuclei yield ratios for *A*=3 and *A*=7 for different reaction systems. The slope values in the last row were obtained from the fitting of the ordinate values of Fig. 2 (a), after averaging over different reaction systems as shown in Fig. 2 (b).

| Reaction | Slope parameter (*a′*) | |
|---|---|---|
| | *A*=3 | *A*=7 |
| $^{78}$Kr+$^{58}$Ni | 6.83±0.17 | 6.70±0.41 |
| $^{78}$Kr+$^{64}$Ni | 7.27±0.22 | 6.92±0.65 |
| $^{86}$Kr+$^{58}$Ni | 6.74±0.24 | 6.93±0.72 |
| $^{86}$Kr+$^{64}$Ni | 6.59±0.22 | 7.00±0.61 |
| Mean[a] | 6.90±0.17 | 6.87±0.31 |

[a] obtained by fitting the data averaged over different reaction systems

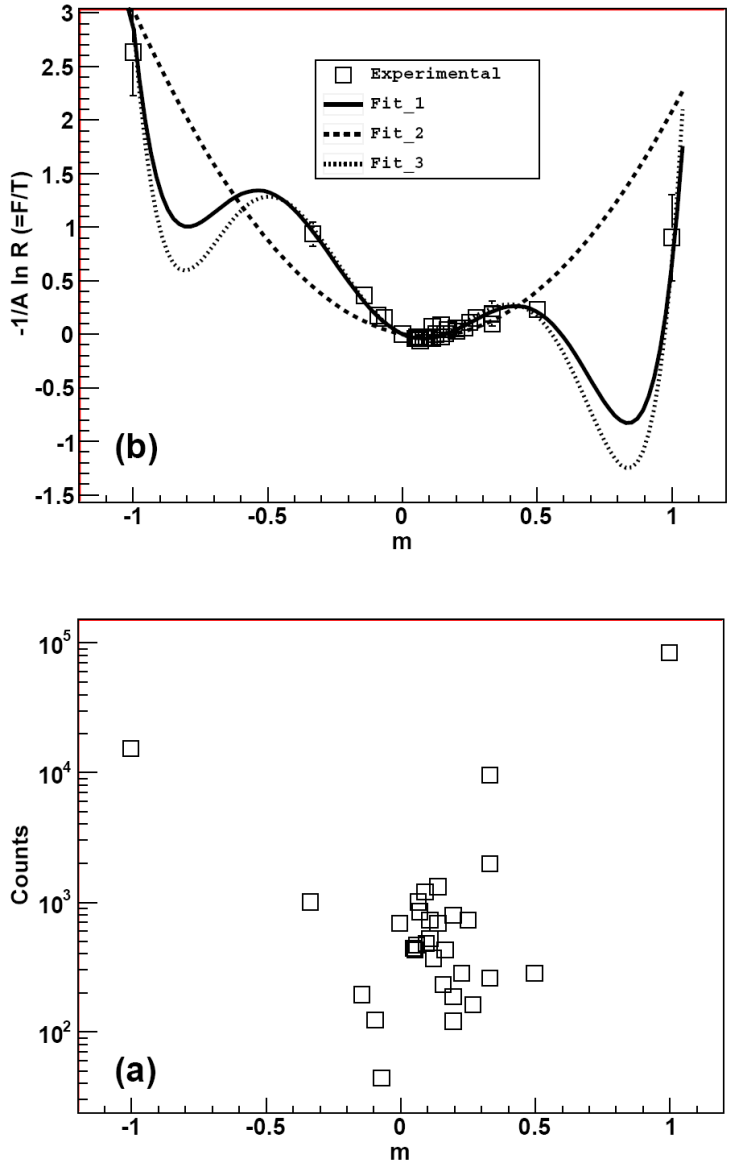

Fig. 1. (a) (bottom panel) Plot of fragment yields arising from the fragmentation of the quasiprojectiles with isospin asymmetry of 0.169 as a function of their isospin asymmetry *m*

(b) (Top panel) Plot of fragment free energies, calculate by the procedure discussed in ref. [11], as a function of their isospin asymmetry *m*. Solid line (Fit_1) is fit to data with Landau equation (Eq (1)). Fit_2 is a fit to the data using Eq (1) with only first and last term. Fit_3 is a fit to the data using '*H/T*' and parameter '*a*' obtained from the analysis of mirror nuclei yield ratio data

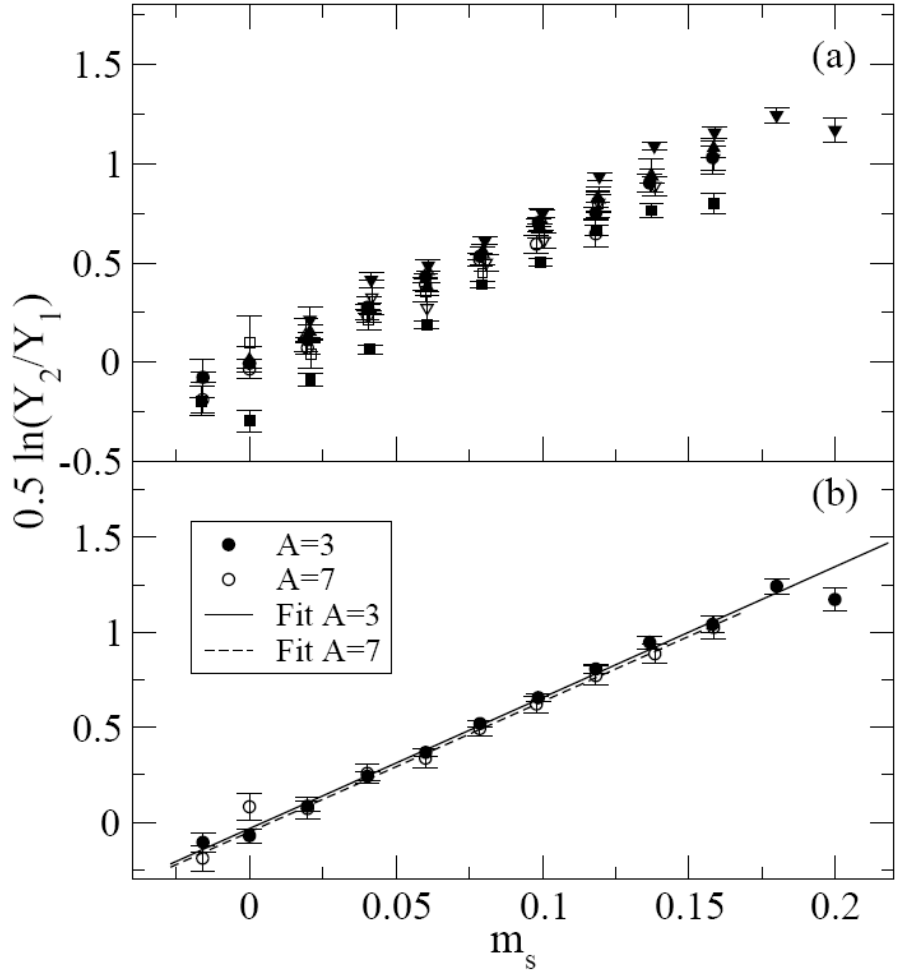

Fig. 2.  (a) Plot of '$0.5\ln(Y_2/Y_1)$' for $A=3$ and 7 as a function isospin asymmetry ($m_s$) of the quasiprojectile source for the reactions $^{78,86}$Kr+$^{86,64}$Ni at beam energy of 35 MeV/nucleon. Cirlcle, square, up triangle and down triangle correspond to $^{78}$Kr+$^{58}$Ni, $^{78}$Kr+$^{64}$Ni, $^{86}$Kr+$^{58}$Ni and $^{86}$Kr+$^{64}$Ni reactions respectively. Filled and Open symbols correspond to $A=3$ and $A=7$ respectively. The subscripts '1' and '2' refer to the neutron deficient and neutron rich members of the mirror nuclei pair.

(b) Plot of '$0.5\ln(Y_2/Y_1)$', averaged over different reaction systems for $A=3$ and 7 as a function of $m_s$. Solid and dashed lines are linear fit to the data for $A=3$ and 7 respectively.

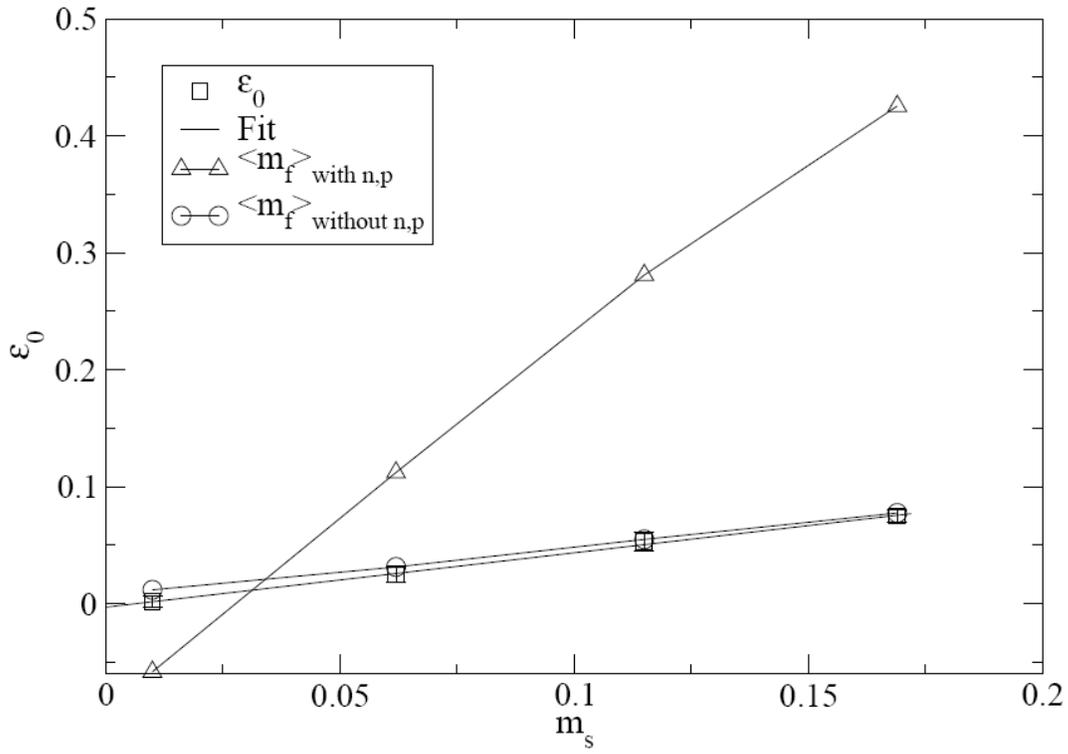

Fig. 3. Plot of the position of the central minima in free energy ($\varepsilon_0$) as a function of isospin asymmetry of the source ($m_s$). Average isospin asymmetry of the fragments $\langle m_f \rangle$ calculated with and without neutrons and protons are also shown in the figure for comparison.

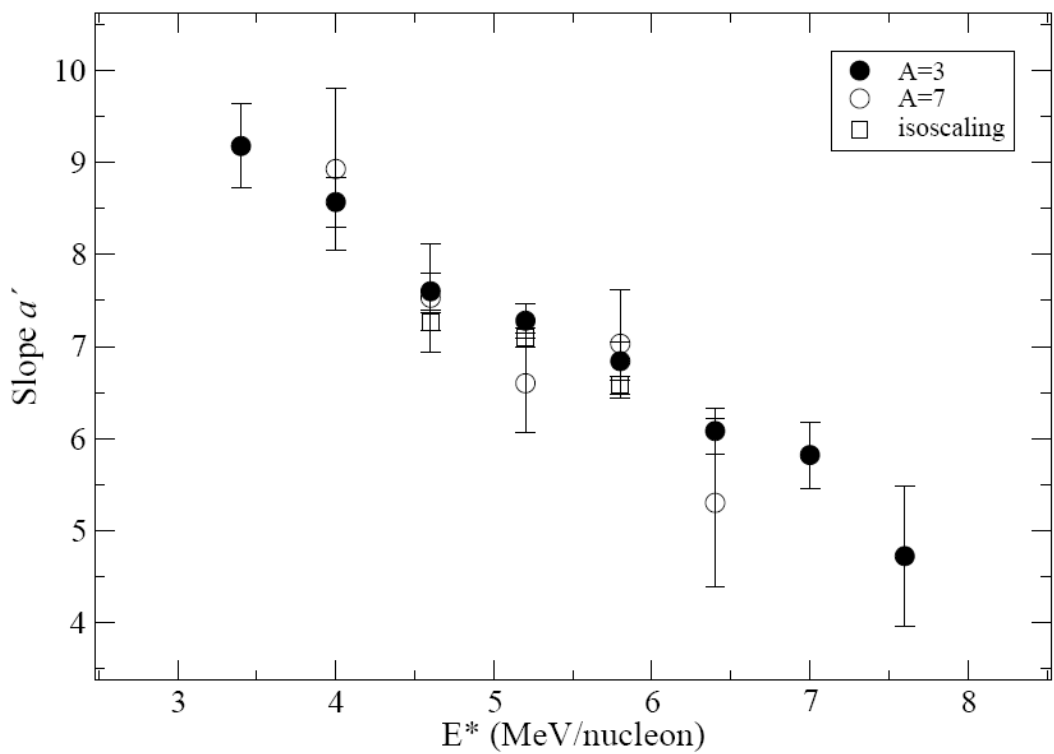

Fig. 4. Slope ($a'$) values, obtained by fitting the plots of mirror nuclei yield ratios as a function of isospin asymmetry of the source for $A$=3 (filled circle) and 7 (open circle), as a function of excitation energy of the quasi projectile. Squares were obtained by fitting the isoscaling plots, similar to that in Fig. 5 with a gate on excitation energy.

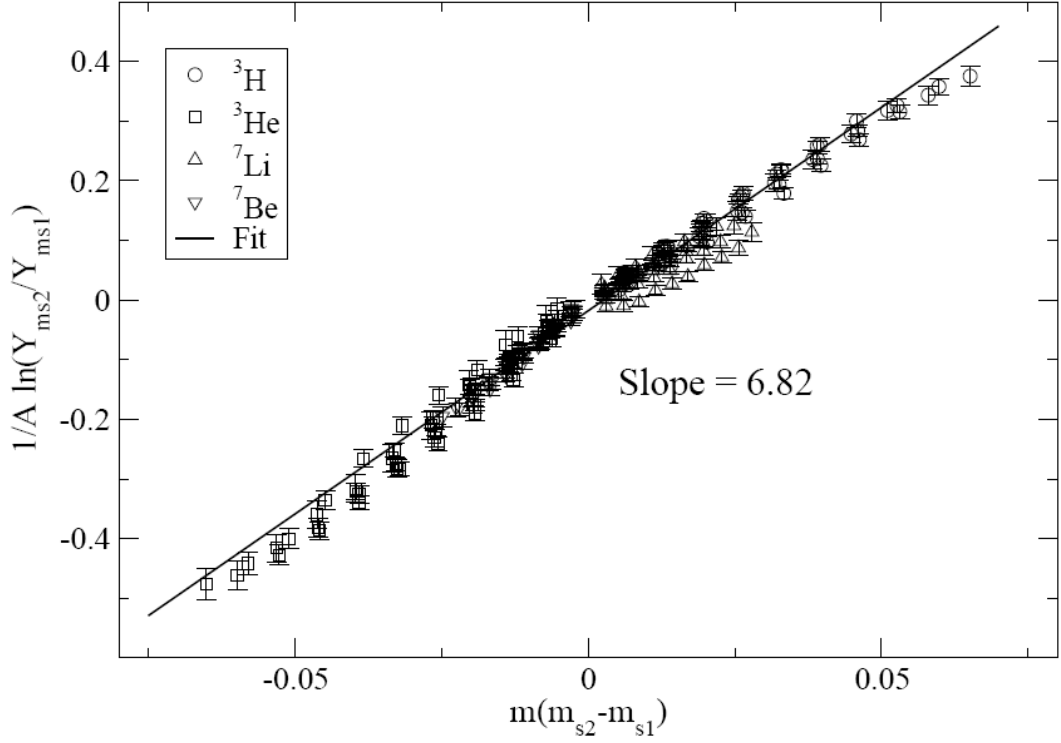

Fig. 5. Plot of '$1/A \ln(Y_{ms2}/Y_{ms1})$' as a function of $m(m_{s2}-m_{s1})$. $Y_{ms1}$ and $Y_{ms2}$ are, respectively, the yields of a fragment from sources with isospin asymmetry of $m_{s1}$ and $m_{s2}$. $m$ is the isospin asymmetry of the fragment. $Y_{ms}$ was normalized with respect to the total number of events in the bin corresponding to $m_s$. Yield ratios were calculated for various possible combinations of $m_s$ bins such that $m_{s2}>m_{s1}$.